# Interlayer Exciton Optoelectronics in a 2D Heterostructure p-n Junction

Jason S. Ross[1], Pasqual Rivera[2], John Schaibley[2], Eric Lee-Wong[2], Hongyi Yu[3], Takashi Taniguchi[4], Kenji Watanabe[4], Jiaqiang Yan[5,6], David Mandrus[5,6,7], David Cobden[2], Wang Yao[3], Xiaodong Xu[1,2,*]

Affiliations:
[1]Department of Materials Science and Engineering, University of Washington, Seattle, Washington 98195, USA
[2]Department of Physics, University of Washington, Seattle, Washington 98195, USA
[3]Department of Physics and Center of Theoretical and Computational Physics, University of Hong Kong, Hong Kong, China
[4]Advanced Materials Laboratory, National Institute for Materials Science, Tsukuba, Ibaraki 305-0044, Japan
[5]Materials Science and Technology Division, Oak Ridge National Laboratory, Oak Ridge, Tennessee, 37831, USA
[6]Department of Materials Science and Engineering, University of Tennessee, Knoxville, Tennessee, 37996, USA
[7]Department of Physics and Astronomy, University of Tennessee, Knoxville, Tennessee, 37996, USA

*Correspondence to: xuxd@uw.edu

**Abstract:** Semiconductor heterostructures are backbones for solid state based optoelectronic devices. Recent advances in assembly techniques for van der Waals heterostructures has enabled the band engineering of semiconductor heterojunctions for atomically thin optoelectronic devices. In two-dimensional heterostructures with type II band alignment, interlayer excitons, where Coulomb-bound electrons and holes are confined to opposite layers, have shown promising properties for novel excitonic devices, including a large binding energy, micron-scale in-plane drift-diffusion, and long population and valley polarization lifetime. Here, we demonstrate interlayer exciton optoelectronics based on electrostatically defined lateral p-n junctions in a $MoSe_2$-$WSe_2$ heterobilayer. Applying a forward bias enables the first observation of electroluminescence from interlayer excitons. At zero bias, the p-n junction functions as a highly sensitive photodetector, where the wavelength-dependent photocurrent measurement allows the direct observation of resonant optical excitation of the interlayer exciton. The resulting photocurrent amplitude from the interlayer exciton is about 200 times smaller compared to the resonant excitation of intralayer exciton. This implies that the interlayer exciton oscillator strength is two orders of magnitude smaller than that of the intralayer exciton due to the spatial separation of electron and hole to opposite layers. These results lay the foundation for exploiting the interlayer exciton in future 2D heterostructure optoelectronic devices.

## MAIN TEXT

The diverse family of van der Waals materials offers a platform for building heterojunctions with designer functionalities at the atomically thin limit[1,2]. Two-dimensional (2D) materials with metallic (graphene), semiconducting (group VIB transition metal dichalcogenides), and insulating (boron nitride, hBN) properties can be stacked into arbitrarily complex heterostructures (HSs). This has led to exciting progress both in science, such as the artificial superlattice yielding Hofstadter's butterfly[3-5], and bandgap engineered devices, in particular efficient light emitting diodes and photodetectors[6,7].

In individual layers of transition metal dichalcogenides (TMDs) the direct bandgap[8,9] enables diverse applications in photovoltaics and light emitting devices (LEDs)[10-16], while the valley-contrasting physics and resultant optical selection rules[17-21] suggests a new paradigm of optoelectronics utilizing the valley pseudospin[22]. Vertically stacking two different TMDs forms a new type of optically active heterojunction with type-II band alignment[23-26]. This results in a built-in vertical p-n junction[6,27,28] at the atomic scale and causes optically excited electrons and holes to be subsequently separated into opposite layers[29,30].

Due to the small interlayer separation, the spatially separated electron and hole still experience strong Coulomb interaction and thus form a tightly bound interlayer exciton ($X_I$)[30-32]. $X_I$ inherits valley-dependent properties from monolayers[33], and exhibits an electrically tunable population lifetime of up to microseconds and a valley lifetime of tens of ns[34]. However, the role of $X_I$ in optoelectronic devices, such as light emitting diodes (LEDs) and photodetectors, remains elusive due to the following facts. As a result of the reduced overlap of the spatially separated electron and hole wave functions, the $X_I$ oscillator strength is expected to be dramatically reduced compared to that of intralayer excitons in individual monolayers. In addition, the inevitable twist between the monolayers during the fabrication causes misalignment of the band edges in momentum space, i.e., $X_I$ is momentum-indirect in its ground state[33,34]. Further, $X_I$ has only been observed so far in photoluminescence (PL) measurements. The weak oscillator strength and momentum-indirect nature make it challenging to directly observe by resonant optical excitation. A means to resonantly probe $X_I$ and thereby determine its oscillator strength is necessary for the observation of such phenomena as interlayer exciton valley currents[33], and for the development of optoelectronic devices exploiting interlayer excitons.

In this paper, we report electrostatically defined lateral p-n junction devices on $MoSe_2$-$WSe_2$ heterobilayers which enables us to resonantly probe the optoelectronic response of $X_I$. The p-n junction design is adapted from Ref. 10. Multiple palladium back gates separated by 300 nm gaps are used to induce adjacent p and n regions above them (Figs. 1a-c). The following layers are stacked on the gates using a modified version of the polycarbonate-based dry transfer technique[35]: first, 10 nm hBN is placed on the palladium back gates as the gate dielectric; next, overlapping monolayers of $MoSe_2$ and $WSe_2$; and last, a capping monolayer of BN which serves both to protect the TMDs and as a thin dielectric layer for tunneling contacts[36] (t-BN in Fig. 1c). Gold source and drain contacts are defined directly on top of the tunneling BN layer. The crystal axes of the $MoSe_2$ and $WSe_2$ are identified for alignment by polarization-resolved second harmonic generation measurements[37-39]. Minimizing the twist angle aligns the Brillouin zones in both layers to minimize the momentum-indirect effect and thereby facilitate the interlayer exciton emission[33,34] (see Supplemental Figure S1).

Figure 1b is a zoom on the optical image of the active area of the device in Fig. 1a, with blue and green dashed lines added to indicate the boundaries of the monolayer $WSe_2$ and $MoSe_2$, respectively. There are three lateral p-n junctions in this particular device, one labeled W1 in the monolayer $WSe_2$, one labeled HS1 in the heterobilayer, and one labeled HS2 where the edge of the $MoSe_2$ monolayer lies in the gap between the two gates. Below we focus on results from HS2, while results from HS1 and W1 are shown and compared in the Supplemental Figure S2. In HS2, the right side of the gap is heterobilayer while the left side is monolayer $WSe_2$, as indicated in Fig. 1c. Figure 1d shows a sketch of the ungated band edges. The kink of the $WSe_2$ bands at the junction indicates the modification of $WSe_2$ electronic structure at the heterobilayer due to the type II band alignment[27,29,31,34], where $MoSe_2$ has the lower conduction band, favoring electron injection, while $WSe_2$ has the higher valence band, favoring hole injection. Importantly, the lateral p-n junction, with contact to each constituent layer of HS2, allows for layer specific carrier injection of the type preferred according to the HS band alignment and the directional interlayer charge transfer.

The black curve in Figure 1e shows the ambipolar transport via sweeping global back-gate doping ($V_g$ in Fig. 1d), which identifies hole conduction turns on at $-2$ V while electron conduction turns on at $+4$ V at a bias voltage of $V_{SD} = 500$ mV. The source-drain current-voltage (red), under p-doped conditions ($V_g = -2.5$ V), also resembles those seen in earlier lateral monolayer TMD LEDs[10]. Setting the back gates to opposite voltages establishes a lateral p-n junction, as shown in Figure 1f. With $V_{BG1} = -1$ V and $V_{BG2} = 5$ V, sweeping the source-drain bias $V_{SD}$ from $-3.5$ to $3.5$ V shows the expected diode behavior (Fig. 1g; see Supplemental Note 1 for explanation of device voltages).

The formation of p-n junction allows us to investigate electroluminescence by applying forward bias. Figure 2a shows a microscope image of the device and a corresponding image of the integrated EL (red) at $V_{SD} = 3.5$ V overlaid on the device (black and white reflection image). EL is seen from the heterobilayer region in the gap between the gates (red arrow) but not the $WSe_2$-only region (black arrow). The reason for this is apparent on comparing the spectrum of the EL with the zero-gate PL under 532 nm excitation at 5 K (Fig. 2b). In the PL (black) we see emission from the intralayer A exciton manifolds, which include overlapping contributions from neutral, charged, and localized excitons in both $MoSe_2$ and $WSe_2$ monolayers (1.56 to 1.74 eV)[20,40]. The dominant emission below 1.4 eV is characteristic of the interlayer exciton[31]. However the EL (red) is dominated by the interlayer emission with only a small intralayer signal at ~1.62 eV, probably from the $MoSe_2$ trion[40]. Hence, despite the small electron-hole wavefunction overlap in $X_I$, having lower energy than the intralayer excitons makes it the dominant radiative recombination pathway at the junction (see cartoon inset in Fig. 2b). In other words, most of the electrons injected in the $MoSe_2$ conduction band and the holes in the $WSe_2$ valence band meet in the junction, bind into $X_I$ due to strong Coulomb interactions, and then recombine to emit light.

Fig. 2c shows how the low energy EL spectrum varies with $V_{SD}$ at 30 K. The peak interlayer emission shifts from 1.355 eV to 1.38 eV as $V_{SD}$ varies from 2.3 V to 3.5 V. The shift resembles the effect seen in previous monolayer TMD LEDs[10], where higher-energy states are occupied more at larger source-drain bias. However, in the present case a vertical electric field can also alter the energy of $X_I$ energy, via the Stark effect, by as much as 40 meV[31]. This could contribute to the shift since biasing the source relative to the back gate changes the vertical field. Future work is therefore needed to separate the effects of state filling and Stark shift in the energy of $X_I$.

Now we turn to the determination of interlayer exciton oscillator strength. To do this, HS2 was operated as a photodetector with $V_{BG1} = -3$, $V_{BG2} = +3$, and $V_{SD} = 0$ and resonant photocurrent (PC) measurements were performed. Figure 3a is sketch of the band diagram in this configuration, where photoexcited electron-hole pairs are separated by the applied electric field at the p-n junction, generating current. The inset of Figure 3b shows a scanning PC map (red intensity indicates current) under 2 µW, 632 nm excitation overlaid on a laser reflection map (black and white), which shows the PC response is localized to the p-n junction region. The main panel shows the PC amplitude (log scale) as a function of laser photon energy. The resonances between 1.6 and 1.7 eV are characteristic of the A excitons and trions in monolayers of $WSe_2$ and $MoSe_2$[20,40], and the PC amplitude drops drastically once the photo-excitation is below the lowest energy bright intralayer exciton, which is the $MoSe_2$ trion around 1.63 eV. Remarkably, there is also a peak at about 1.41 eV, corresponding to $X_I$. This is the first observation of resonant optical excitation of an interlayer exciton.

The peak energy of the $X_I$ in PC spectrum is about 20-30 meV higher than both interlayer EL and PL peaks. However, gate dependent PL measurements show that the Stark tuning of the $X_I$ PL cannot fully account for the blue shift of the peak in the wavelength dependent PC measurements (see Supplemental Figure S3). Instead, this can be understood by considering that with finite twisting angle the direct optical transitions for interlayer exciton states are at finite kinematic momenta[33]. These points are the $X_I$ light cones in the dispersion diagram, where direct interconversion with photon conserves momentum, without the need for scattering. This unique aspect of the interlayer exciton dispersion is shown in Fig. 3c, where resonant excitation (orange squiggly line in figure 3b) generates interlayer excitons with finite kinetic energy. In contrast, EL and PL emissions (red squiggly line in figure 3c) have contributions from cold excitons near the $X_I$ dispersion minimum, which can emit photon through scattering processes, e.g. phonon-assisted recombination. Due to the large binding energy and long lifetime of $X_I$, their density should peak near the dispersion minimum in the PL and EL measurements. Significant low energy emissions from these cold $X_I$ red shift their spectra relative to the PC, which is primarily sensitive to $X_I$ at the light cone. In device HS2, polarization-resolved second harmonic generation measurements confirm the twist angle to be about 3 degrees (Supplemental Figure S4), which corresponds to an exciton kinetic energy of tens of meV at the light cones[33]. Therefore, the interlayer PC resonance is expected to be at higher energy than the PL and EL, which is consistent with our measurement results.

Resonant excitation of the $X_I$ enables us to compare the interlayer and intralayer exciton oscillator strengths. To do so, the spot size and power of the laser vs. wavelength was calibrated using a knife edge and photodiode, respectively. The PC could then be normalized to the laser irradiance, giving photoresponse units of mA/(µW/µm$^2$). In the intralayer region (Fig. 4a) we see strongly coupled (large photoresponse) features, as described above. In contrast, the interlayer photoresponse (Figure 4b) is about 200 times weaker. If we assume the photocurrent is proportional to the number of optically generated excitons, which has a linear dependence on exciton oscillator strengths for a given photo-excitation density, then the $X_I$ oscillator strength is two orders of magnitude smaller than for the intralayer excitons. This is consistent with theoretical predictions[33]. Compared to the intralayer exciton, where the electron and hole are confined to the same 2D layer (see illustration in Figure 4c), the interlayer exciton (Figure 4d) has the electron and hole constituent separated to different layers, resulting in reduced dipole matrix element[33]. In addition, the interlayer exciton oscillator strength depends on interlayer separation, which can be sample dependent, and may be further impacted by substrate interactions, such as with the BN encapsulating layers. We expect that the demonstrated resonant excitation of interlayer excitons will provide a way to optically generate valley current using linearly polarized light, as proposed by Ref. 32.

**Methods:**

The PC measurements were performed using actively power-stabilized continuous wave output from a frequency tunable and narrow band ($< 50$ kHz) Ti:Sapphire laser (MSquared SolsTiS). The laser output was passed through an acousto-optic modulator (AOM), and the first order diffracted beam was used as the excitation source. A beam sampler reflected a small fraction of the first order beam to measure its intensity on a silicon photodiode. This provided the input signal to a high-speed servo controller (New Focus LB1005) which used PID control over the AOM to lock the absolute laser power. At each wavelength, the setpoint of the servo system was adjusted to achieve the desired laser intensity, measured just before the objective (Olympus LCPLN50XIR) using a wavelength calibrated power meter (Thorlabs PM100D). This home-built system provides RMS fluctuations of the laser power of less than 1% across the entire range used in the PC measurement.

**Acknowledgements:**
This work is supported by AFOSR (FA9550-14-1-0277) and NSF EFRI-1433496. K.W. and T.T. acknowledge support from the Elemental Strategy Initiative conducted by the MEXT, Japan and a Grant-in-Aid for Scientific Research on Innovative Areas "Science of Atomic Layers" from JSPS. JY, DM were supported by US DoE, BES, Materials Sciences and Engineering Division. H.Y. and W.Y. were supported by the Croucher Foundation (Croucher Innovation Award), and the Research Grants Council and University Grants Committee of Hong Kong (HKU17305914P, HKU9/CRF/13G, AoE/P-04/08). X.X. acknowledges a Cottrell Scholar Award, Boeing Distinguished Professorship in Physics, and support from the State of Washington–funded Clean Energy Institute.

**Author Contribution**:
XX conceived the project. JSR fabricated the devices and performed the measurements, assisted by PR, JS, and ELW. JSR, HY, XX and WY analyzed the data. JY and DGM provided and characterized the bulk $MoSe_2$ and $WSe_2$ crystals. TT and KW provided bulk BN crystals. JSR, XX, PR, WY, HY, and DHC wrote the paper. All authors discussed the results.

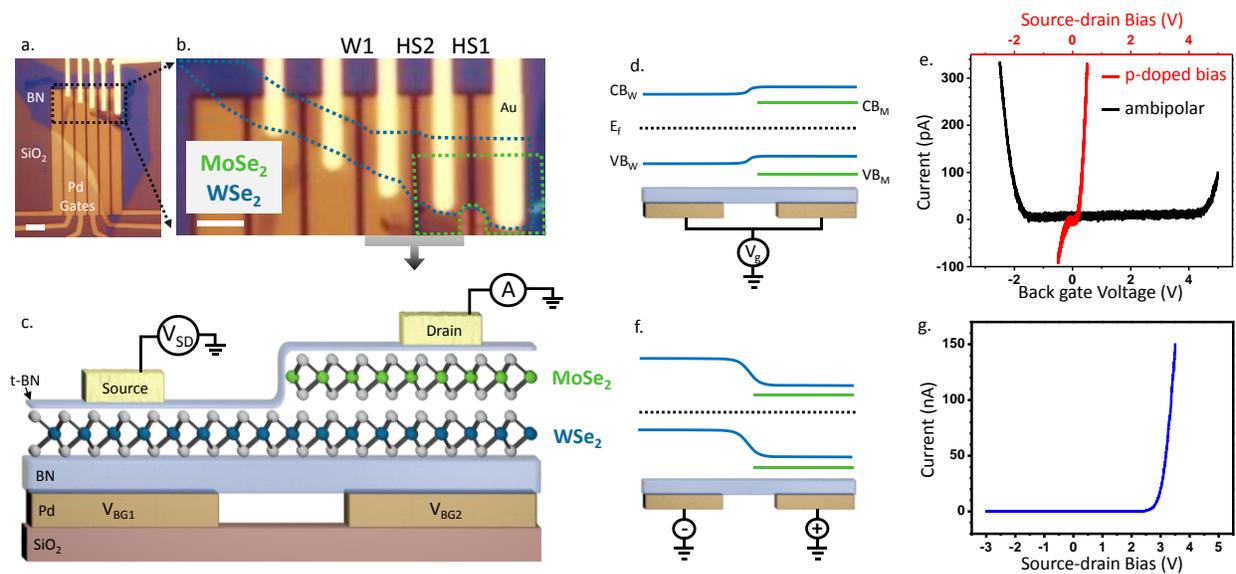

**Figure 1| Device description and electrical transport. a.** Wide view microscope image of device. Scale bar 2 μm. **b.** Zoom on active area of showing three labeled junctions. Dotted lines outline MoSe$_2$ (green) and WSe$_2$ (blue) monolayers. Scale bar 2 μm. **c.** Cartoon of device HS2 cross-section showing the arrangement of the layers and electrical contacts. **d.** Band diagram across HS2 with zero gate-doping showing conduction band (CB) and valence band (VB) for WSe$_2$ (blue) and MoSe$_2$ (green). Dotted line is the Fermi level (E$_F$). Here we show both gates are tied together to provide the same doping on both sides of the junction. **e.** An ambipolar gate dependence is seen in this configuration (black curve with V$_{SD}$ = 500 mV). The red trace is an I-V curve with p-doping at V$_g$ = -2.5 V. **f.** Band diagram across HS2 with the gates separately biased with opposite polarity to form a p-n junction. **g.** I-V curve in this configuration, showing expected diode behavior.

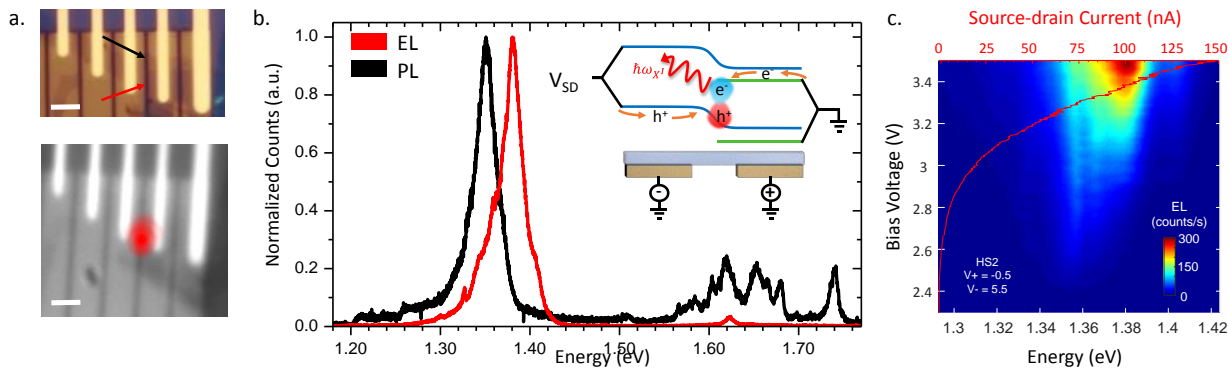

**Figure 2| Interlayer exciton electroluminescence. a.** Top: microscope image of device area with arrows pointing to WSe$_2$-only region (black) and HS region (red) of HS2 device. Bottom: electroluminescence (EL) image (red) overlaid on top of white light illuminated spectrometer image of device area (black and white). Scale bars 2 μm. **b.** Spectra of EL (red) and photoluminescence (black) for HS2. Inset cartoon illustrates which layer each source and drain contact (black lines) is connected to, how carriers are injected into the device, and the p-n junction recombination area (orange arrows). In the center, the carriers bind into an interlayer exciton before emitting light. **c.** Source-drain bias dependence of interlayer-exciton EL with corresponding I-V trace (red) overlaid.

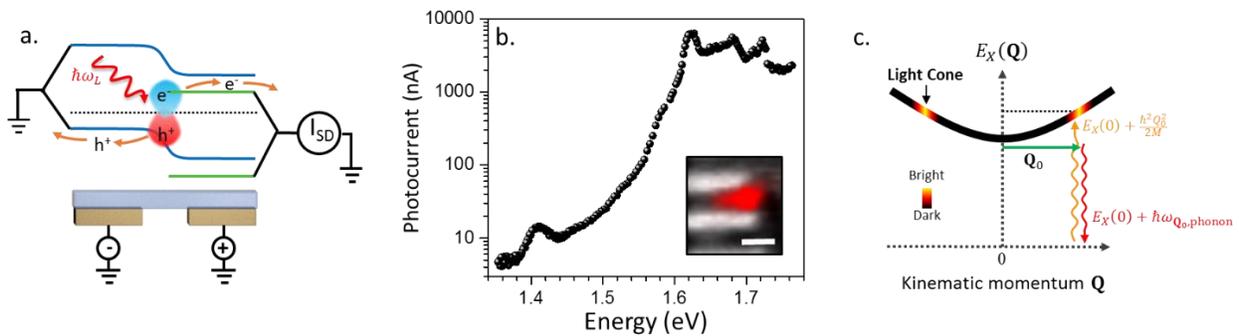

**Figure 3| Direct light coupling of interlayer exciton. a.** Photocurrent (PC) $I_{SD}$ is generated at $V_{SD}=0$ where incident laser light creates electron-hole pairs that are separated by the p-n junction field and collected by the contacts. **b.** Dependence of PC on laser photon energy. Inset: scanning PC map (red intensity is current amplitude) overlaid on a corresponding scanning reflection map (black and white) of the junction. Scale bar is 2 μm. **c.** Interlayer exciton energy-momentum diagram. The dispersion crosses the light cone (yellow) at finite kinetic momentum ($Q_O$). Excitons at the bottom of the dispersion curve (with energy $E_X(0)$) can be scattered to the light cone to emit light (red transition) assisted by a phonon. Direct optical excitation (orange transition) creates interlayer exciton with finite kinetic energy and produces PC by subsequent electron-hole separation.

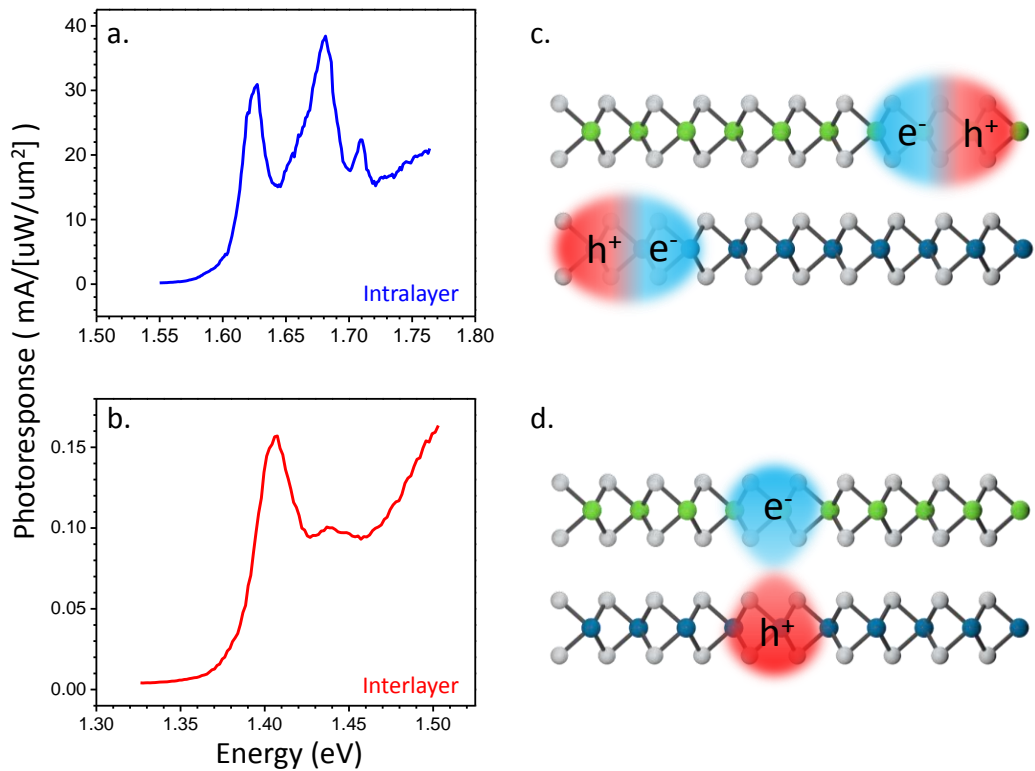

**Figure 4| Photoresponse comparison of interlayer to intralayer excitons. a** and **b,** show photoresponse as a function of laser excitation energy for the intralayer and interlayer energy ranges, respectively. **c** and **d,** depict wave function overlap between electrons and holes for intralayer and interlayer excitons, respectively. The interlayer exciton has smaller electron-hole wave function overlap, and thus smaller dipole matrix element and oscillator strength, compared to the intralayer exciton.

# Supplemental Materials for

# Interlayer Exciton Optoelectronics in a 2D Heterostructure p-n Junction

Jason S. Ross[1], Pasqual Rivera[2], John Schaibley[2], Eric Lee-Wong[2], Hongyi Yu[3], Takashi Taniguchi[4], Kenji Watanabe[4], Jiaqiang Yan[5,6], David Mandrus[5,6,7], David Cobden[2], Wang Yao[3], Xiaodong Xu[1,2,*]

Affiliations:
[1]Department of Materials Science and Engineering, University of Washington, Seattle, Washington 98195, USA
[2]Department of Physics, University of Washington, Seattle, Washington 98195, USA
[3]Department of Physics and Center of Theoretical and Computational Physics, University of Hong Kong, Hong Kong, China
[4]Advanced Materials Laboratory, National Institute for Materials Science, Tsukuba, Ibaraki 305-0044, Japan
[5]Materials Science and Technology Division, Oak Ridge National Laboratory, Oak Ridge, Tennessee, 37831, USA
[6]Department of Materials Science and Engineering, University of Tennessee, Knoxville, Tennessee, 37996, USA
[7]Department of Physics and Astronomy, University of Tennessee, Knoxville, Tennessee, 37996, USA

*Correspondence to: xuxd@uw.edu

## Supplemental Discussion S1. Description of voltages and device function

The reason for the smaller voltage on $V_{BG1}$ is because of the interplay with $V_{SD}$, which is applied to the source contact electrode on WSe$_2$, whereas the drain electrode on MoSe$_2$ is grounded. As $V_{SD}$ is ramped up and drives source-drain current, the voltage drop across the boron nitride (BN) dielectric increases and thus the WSe$_2$ hole doping increases such that at $V_{SD} = 3.5$ V, the gating in WSe$_2$ is comparable to the gating in MoSe$_2$ when $V_{BG2} = 5$ V. If the gate voltages were equal and opposite, say -5 and +5 V, as $V_{SD}$ is increased the BN dielectric would break down before forward bias conduction occurs. This is a limitation of these electrostatically defined lateral p-n junctions. In the negative $V_{SD}$ regime, we have swept $V_{BG1}$ to $-8$ V in order to ensure comparable p and n doping on respective sides of the device and no reverse bias conduction is observed, confirming we have a diode.

**Supplemental Figure S1. Heterostructure twist angle and interlayer exciton light cones.**

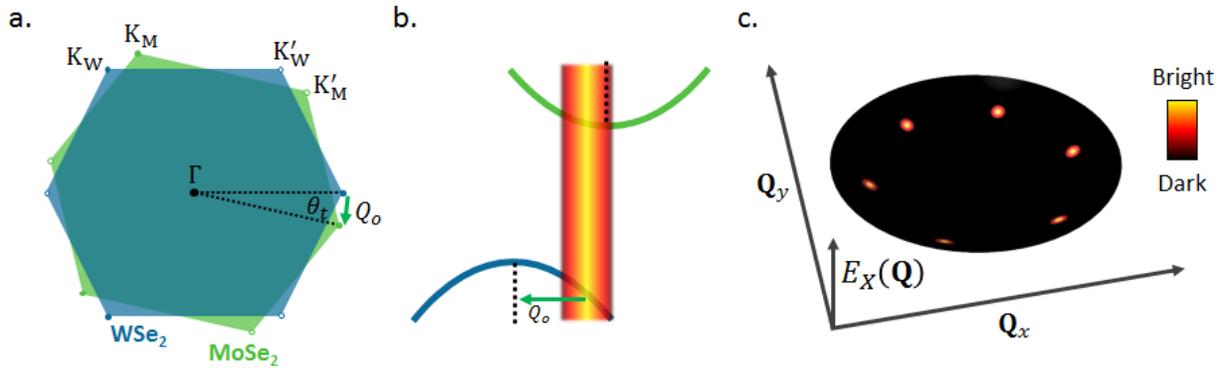

**Figure S1 | Heterostructure twist angle and interlayer exciton light cones. a.** A small twist angle ($\theta_t$) results in the K valleys of constituent materials becoming momentum mismatched in the first Brillouin zone. The displacement vector $\mathbf{Q_o}$ (green arrow) is the combined momentum of an electron and hole necessary to recombine for an interlayer transition[1]. **b.** This means there's a light cone at $\mathbf{Q_o}$ away from the K point of a given material (e.g. the valence band edge of WSe2 in this figure). **c.** Due to six-fold rotation symmetry, there are six light cones in the 2D exciton energy dispersion ($E_X(\mathbf{Q})$) with the interlayer exciton optically dark at $\mathbf{Q} = 0$. *Figures recreated from reference [1].*

**Supplemental Figure S2. Electroluminescence of additional devices**

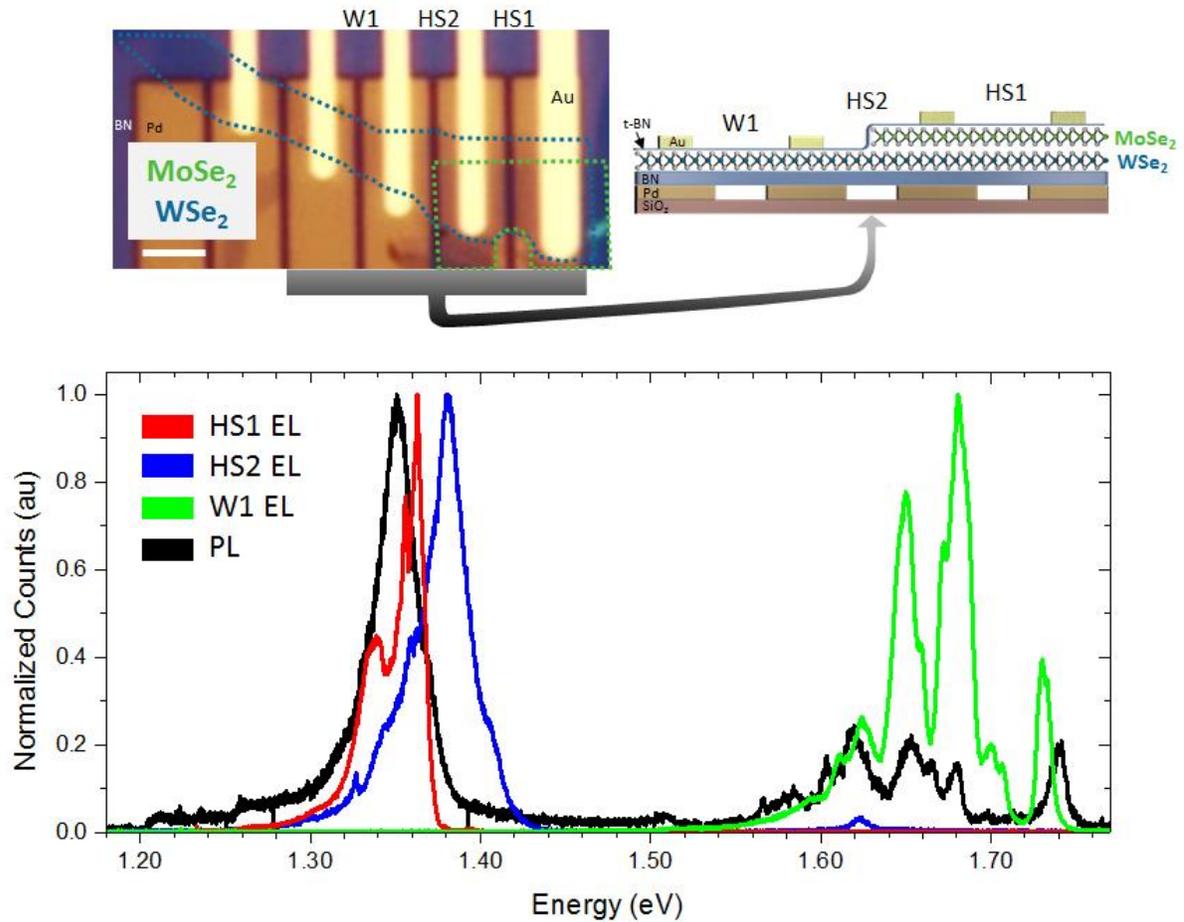

**Figure S2 | Electroluminescence of additional devices.** Electroluminescence (EL) from devices under similar voltages as discussed in the main text are compared above to the photoluminescence (PL) of undoped HS2 (black). In red is the full-HS device HS1, which has less blue shift to the PL than HS2 in blue. These relative interlayer emission energies can all be explained by a combination of the Stark effect and state filling, as discussed in the main text. In practice, HS1 was found to have less Stark shifts than HS2. In the main text Fig. 2c we see that at lower $V_{SD}$ of about 2.5 V, HS2 EL can indeed match the PL at 1.36 eV. Finally, in green we have the emission from the single layer WSe$_2$ device W1 which shows EL only from intralayer exciton states between 1.6 and 1.74 eV[2].

**Supplemental Figure S3. Global backgate dependent photoluminescence from HS2**

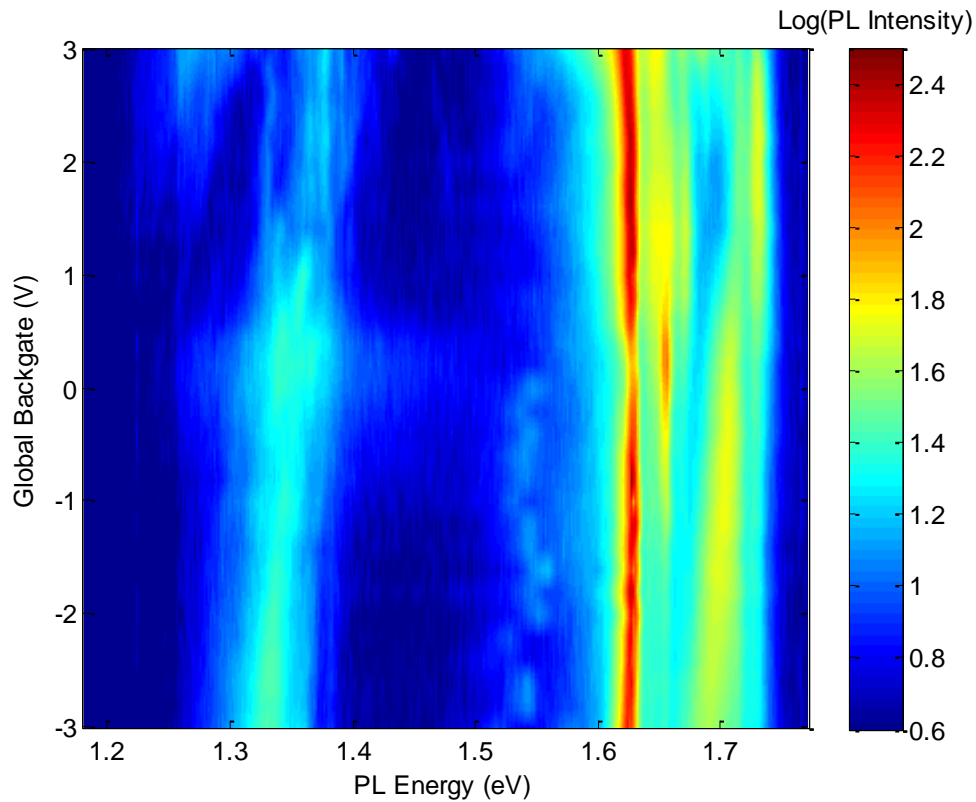

**Figure S3 | Global backgate dependent photoluminescence from HS2.** Photoluminescence (PL) from HS2 at a temperature of 30K under excitation with 2 µW laser power from a He-Ne light source (632 nm). All backgates are changed together while the top contacts remain grounded. The Stark tuning of the interlayer exciton shifts the low energy PL from ~1.33 eV to 1.38 eV. Importantly, the interlayer exciton PL is always below that of the peak in the wavelength dependent photocurrent, shown in Figs. 3b and 4b.

**Supplemental Figure S4. Twist angle characterization**

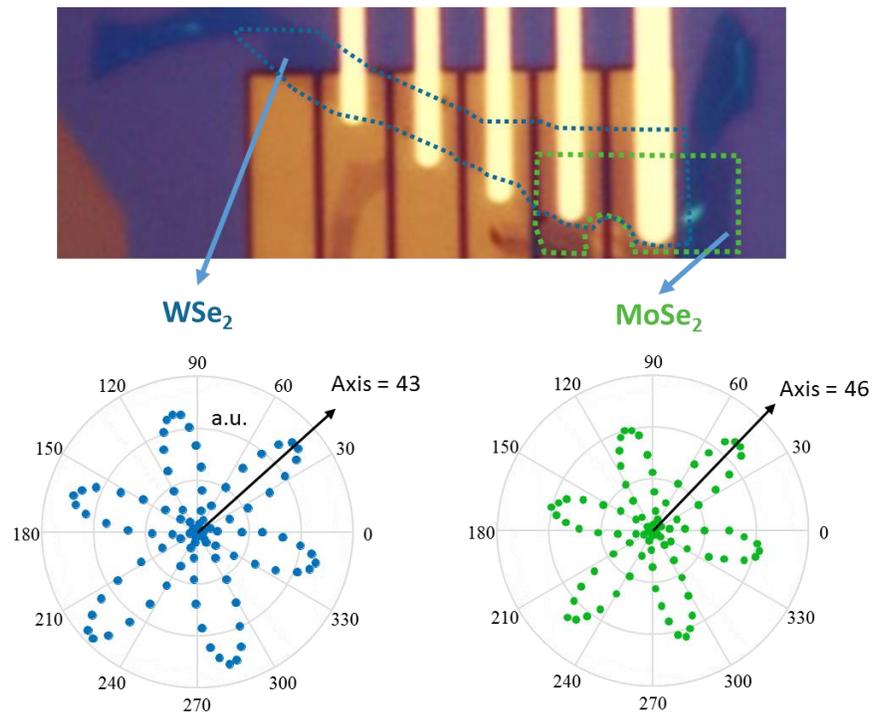

**Figure S4 | Second harmonic generation (SHG) vs. linear polarization angle of incident light.** Data taken from monolayer areas of constituent materials as indicated by the blue arrows. For $WSe_2$ ($MoSe_2$), incident light set to 1480 (1560) nm and SHG signal collected at 740 (780) nm. The maximum SHG signal co-polarized with the linearly-polarized excitation laser corresponds to the armchair axes of each crystal[3-5]. Looking at the difference in the armchair axes of each crystal, we estimate a 3-degree twist angle for the heterostructure region.